\documentclass[sn-nature,iicol]{sn-jnl}

\usepackage{graphicx}%
\usepackage{multirow}%
\usepackage{amsmath,amssymb,amsfonts}%
\usepackage{amsthm}%
\usepackage{mathrsfs}%
\usepackage[title]{appendix}%
\usepackage{xcolor}%
\usepackage{textcomp}%
\usepackage{manyfoot}%
\usepackage{booktabs}%
\usepackage{algorithm}%
\usepackage{algorithmicx}%
\usepackage{algpseudocode}%
\usepackage{listings}%
\usepackage{footnote}
\usepackage{tabularx}
\usepackage[]{mdframed}
\usepackage{framed}

\theoremstyle{thmstyleone}%
\theoremstyle{thmstyletwo}%

\theoremstyle{thmstylethree}%

\begin{document}

\title[Sun as a cosmic ray TeVatron]{Sun as a cosmic ray TeVatron}


\author[1]{\fnm{Prabir} \sur{Banik}}\email{pbanik74@yahoo.com}

\author[2]{\fnm{Arunava} \sur{Bhadra}}\email{aru\_bhadra@yahoo.com}

\author*[1]{\fnm{Sanjay K.} \sur{Ghosh}}\email{sanjay@jcbose.ac.in}

\affil*[1]{Department of Physics, $\&$ Center for Astroparticle Physics $\&$ Space Science,Bose Institute, EN-80, Sector-5, Bidhan Nagar,  Kolkata-700091, India}

\affil[2]{High Energy $\&$ Cosmic Ray Research Centre, University of North Bengal, Siliguri, West Bengal, India 734013}



\abstract{Very recently, HAWC observatory discovered the high-energy gamma ray emission from the solar disk during the quiescent stage of the Sun, extending the Fermi-LAT detection of intense, hard emission between $0.1 - 200$ GeV to TeV energies. The flux of these observed gamma-rays is significantly higher than that theoretically expected from hadronic interactions of galactic cosmic rays with the solar atmosphere. More importantly, spectral slope of Fermi and HAWC observed gamma ray energy spectra differ significantly from that of galactic cosmic rays casting doubt on the prevailing galactic cosmic ray ancestry model of solar disk gamma rays. In this work, we argue that the quiet Sun can accelerate cosmic rays to TeV energies with an appropriate flux level in the solar chromosphere, as the solar chromosphere in its quiet state probably possesses the required characteristics to accelerate cosmic rays to TeV energies. Consequently, the mystery of the origin of observed gamma rays from the solar disk can be resolved consistently through the hadronic interaction of these cosmic rays with solar matter above the photosphere in a quiet state. The upcoming IceCube-Gen2 detector should be able to validate the proposed model in future through observation of TeV muon neutrino flux from the solar disk. The proposed idea should have major implications on the origin of galactic cosmic rays.}



\maketitle

\section{Introduction}\label{sec1}

The High Altitude Water Cherenkov (HAWC) observatory recently discovered a TeV gamma-ray flux from the solar disk ($6.3\sigma$) in the quiescent stage, based on 6.1 years of data \citep{HAWC22}. The spectrum between $0.5-2.6$ TeV is nicely fitted by the power law with spectral index $\Gamma_{\gamma} = -3.62\pm0.14$ \citep{HAWC22}. These findings extend the Fermi-LAT \citep{Abdo11,Ng16,Linden18} observed GeV (between $0.1-200$ GeV) gamma-ray emission (slope $\approx -2.2$)  from the solar disk to TeV energies. The origin of the stated observed gamma ray emission from solar disk in GeV to TeV energies is not known so far; the prevailing models of gamma ray emission from solar disk cannot explain the mentioned observations. The theoretical explanation of the HAWC and Fermi-LAT observed gamma-ray emission from solar disk is one of the challenges of solar astrophysics at present. 

In 2008, the first evidence of gamma-ray emission from the solar disk and its halo in its quiescent stage was revealed by an analysis of archived data from the Energetic Gamma-ray Experiment Telescope (EGRET) \citep{Orlando08}. The Fermi collaboration confirmed the detection of high-energy gamma rays at $0.1-10$ GeV energy range from the quiescent Sun using the first 1.5 years of data \citep{Abdo11}. Further investigations using Fermi-LAT data suggest a harder spectral slope ($\Gamma \approx -2.2$) upto 200 GeV, without any obvious high energy cutoff \citep{Ng16,Tang18,Linden18}. Additionally, a considerable temporal fluctuation of the solar disk gamma-ray flux in anti-correlation with solar activity was identified in the energy range of $1-10$ GeV, implying that the solar magnetic field might play an important role \citep{Ng16,Tang18}. The gamma-ray flux increases by about a factor of two during solar minimum and then gradually decreases as the Sun approaches solar maximum \citep{Linden22}. There was no substantial energy dependence in the time variability of solar disk emission in the $0.1-10$ GeV energy range \citep{Linden22}. The lack of variation in solar disk emission on shorter time frames suggests that it is not influenced by short-lived magnetic outbursts such as solar flares and coronal mass ejections \citep{Linden22}.

Until now, the prevailing theoretical models \citep{Dolan65,Peterson66,Hudson89,Seckel91} predicted gamma-ray emission from the solar disk during the quiescent stage by assuming the interactions of hadronic galactic cosmic rays with nuclei in the solar atmosphere over the entire disk. Seckel et al. (1991) \citep{Seckel91}, the only detailed theoretical investigation to date, published an estimation of gamma-ray flux from the solar disk by taking into account magnetic reflection and the hadronic interaction of galactic cosmic rays with nuclei in the solar atmosphere across the entire disk. However, the measured solar disk emission flux by Fermi-LAT is nearly a factor of seven larger than that expected by the nominal model of Seckel et al. (1991) \cite{Seckel91}. A full numerical simulation study with the FLUKA code \citep{Ferrari05}, considering the galactic cosmic rays, yields a higher gamma-ray flux from the Sun \citep{Mazziotta20} (which depends on the magnetic field nearby and inside the Sun) than that predicted by the nominal model of Ref.~\cite{Seckel91}. But it is still substantially lower than the observed gamma ray flux, particularly at high energies. The TeV findings by HAWC add further concerns to the puzzle of solar-disk emission. The steeper spectral slope, as seen by HAWC, suggests a break in the gamma-ray energy spectrum  at 400 GeV energy \citep{HAWC22}. 

Therefore, the gamma-ray observations raise question on the possibility of solar disk emission originating from the hadronic interaction of galactic cosmic ray particles with the solar atmosphere. This is also supported by the fact that observed gamma-ray flux by Fermi is harder ($\approx -2.2$) at least up to 200 GeV, and by HAWC is softer ($\approx -3.62$), in contrast to the galactic cosmic ray spectral index below knee ($\approx -2.7$). The reported hard gamma-ray spectral slope is unexpectedly compatible with its origin as if formed by hadronic interaction of shock-accelerated cosmic rays with matter. This motivates us to assume that the Sun might accelerate particles up to a few TeV energies with an appropriate flux level in its quiet state.

In the present paper, we would demonstrate that if the quiet Sun accelerates cosmic rays at least up to a few TeV energy in anti-correlation to solar activity, all the features (except the apparent spectral dip around 30-50 GeV: discussed later) of gamma-ray observations from solar disk can be explained consistently considering the interaction of these accelerated cosmic rays with solar matter above the photosphere. We shall also evaluate the neutrino fluxes from the solar disk under the proposed scenario and explore the possibility of their detection by the IceCube and future neutrino observatories. 

\section{Cosmic Rays from the quiet Sun}
The theory of accelerating solar cosmic rays to relativistic energies still needs to be resolved \cite {Miroshnichenko13}. During solar flare/coronal mass ejection (CME) outbursts, the Sun is usually known to accelerate energetic particles up to only a few (15-30) GeV in the solar corona \citep{Peterson59, Chupp73, Dorman2021}. Early muon detector experiments show that solar cosmic rays may accelerate to $\ge 100$ GeV during massive solar flare/CME events \citep{Miroshnichenko13}. We assume that the (quiet) Sun is also a cosmic ray accelerator with maximum energy around a few TeV. This is the only assumption we have made to explain the stated observations. Such an assumption is not implausible as argued below. 

The observations suggest that charged particles are accelerated in various sites in the Sun \cite{Reames99}. Several models for the acceleration of charged particles in the solar environment have been proposed in the literature. The cause of energetic particle generation in the atmosphere of a star (in the quiet stage) may be similar to those in galactic and intergalactic environments. It is due to the transfer of energy from the macro world to runaway particles in dynamic plasma frozen-in magnetic field \cite{Dorman2021} mainly through the diffusive shock acceleration process. Since shock waves are a prevalent phenomenon in the solar atmosphere, which is supposed to play an essential role in heating the outer layer of the solar atmosphere \citep{Mathur22}, diffusive shock acceleration of particles in the solar atmosphere is not unlikely. In fact, the diffusive shock acceleration (first order Fermi acceleration) may be a viable process for proton acceleration up to TeV energies in the Sun's atmosphere. In such an acceleration mechanism, the charged particles scatter back and forth across the shock front several times by the turbulent magnetic field fluctuations, and in each collision the particles gain energy. The charged particles also can gain energy in interaction with parallel shocks by drifting along the shocks. From the power perspective, shocks in the solar chromosphere has enough power to provide necessary energy to the solar cosmic ray particles as discussed in sec.~\ref{secr}. 

The H \& K resonant lines in singly ionized calcium (Ca II) show grains in the chromosphere and transition regions of the Sun, indicating acoustic-shock-like disturbances moving upward in the solar atmosphere. The typical size of the disturbances is $1''.95$ (1425 KM) \citep{Woger06}. The gyro-radius of TeV protons is smaller than the size of the quiet solar atmosphere. So, protons may remain confined over a long period in the atmosphere and thus may encounter dynamic shocks many times, thereby gaining energy. It is supported by the fact that solar canopy-shaped magnetic fields in the quiet chromosphere restrict the leakage or diffusion of charged particles from the chromosphere into the corona, as indicated by the drastic reduction of particle density from the chromosphere to the transition region and the corona \citep{Li22}. To be considered a possible site of acceleration, the accelerated particles should be maintained within the site during acceleration. As per the Hillas criterion, $ R >E_{max}/B $, where B is the magnetic field, $E_{max}$ is the maximum energy of an accelerated proton and R is the size of the accelerating site. 

For a long time, the quiet Sun was believed to be essentially non-magnetic, because only the photospheric granulation could be detected in continuum images \cite{Rubio19}. However, in polarized light, the quiet Sun exhibits a reticular pattern of powerful kilogauss (kG) fields known as the magnetic network, as well as countless weaker, small-scale flux concentrations in the spaces between them known as the solar internetwork (IN) \cite{Rubio19}. The network outlines supergranular cell (typical dimensions of 30,000 km, the largest convective pattern on the solar surface) borders precisely where horizontal flows become downdrafts \cite{Rubio19}. The internetwork (IN) corresponds closely to the inside of the supergranular cells \cite{Rubio19}. In order to explain the observational Hanle effect, Bueno et al.(2004) \cite{Bueno04} reported that the spatially unresolved hidden turbulent flux between the kG flux tubes and the hidden field in the seemingly empty regions is found to be strong ($\sim 130$ G), but the field polarities are mixed. The presence of such intense magnetic field strength order of 100 G in quiet Sun is now clearly favored by modern findings \cite{Rubio19,Orozco12,Aleman18,Sedik23}. 
For a typical magnetic field of the order of 100 G, the chromosphere, which has a depth of around 2000 km above the solar surface of radius 0.7 million km, is a viable site of shock-accelerated protons with $E_{max}$ about 7 TeV. Even in the photosphere, which spans around 500 km vertically, the maximum energy of accelerated cosmic rays may reach TeV energies for an average magnetic field strength of the order of 100 G. 
Note that, the spectral line of Fe II (645.6 nm) provides signature of the upward moving shocks in the photosphere, resulting from an abrupt slow down of the fast (supersonic) horizontal flow of granular plasma in the direction of the intergranular lane \citep{Rybak04}.

The shock velocity in the solar chromosphere is about two orders smaller than that in a supernova remnant (SNR) \citep{Eklund21,Wentzel67}, but the chromosphere magnetic field \citep{Wiegelmann14} is many order higher than that in SNR. Thus, a maximum energy of a few TeV in the solar chromosphere seems plausible theoretically. Moreover, the cosmic ray induced instabilities may lead to magnetic field amplification \citep{Lucek00}, both in the downstream and upstream, which in turn enhances the maximum energy of cosmic ray particles. Another relevant issue is the radiation energy loss, which should not exceed the energy gain during the acceleration process. In the present scenario, pp interaction is the main energy loss mechanism. Here, the cooling time scale due to pp interaction \citep{Gaisser90} is found to be larger than the acceleration time scale \citep{Murase08} for proton energy is lower than $\sim 15$ TeV. 

The temporal variability of solar disk emission does not show a significant energy dependence, suggesting that magnetic-field effects close to the solar surface may predominantly control the cosmic ray acceleration and, consequently, the gamma-ray flux \citep{Linden22}. The dominant part of the magnetic field of quiet chromosphere (the component III as referred in \cite{Li22}) observed in H$_{\alpha}$, Ca II K, or other emission lines is in anti-phase with the solar cycle \citep{Li22}, which explains the anti-correlation of gamma-ray flux from the disk with solar activity \citep{Linden22}. Component-III's magnetic elements in the Sun's quiet region is primarily distributed at low latitudes around the equatorial plane \citep{Li22}. The cosmic ray acceleration scenario by the magnetic field in the chromosphere is thus also consistent with the recent morphology study \citep{Linden22} of gamma-ray emission from the quiet Sun, which shows that the contribution of gamma rays from the equatorial region of the quiet Sun decreases sharply during active period of the Sun whereas that from the polar region is less impacted.

The production spectrum of such solar accelerated cosmic ray (protons) should follow a power law as $\frac{dn_p}{dE_p} \propto {E_p}^{-\alpha_p}$, where $\alpha_p$ is the spectral index. The spectrum can be normalized to an appropriate cosmic ray luminosity as {\small $L_{p} =  {4 \pi R_{\odot}^2 }\displaystyle\int E_p \frac{dn_p}{dE_p} dE_p$}, where $R_{\odot}$ represents the radius of the Sun. The accelerated protons may interact with the cold (non-relativistic) matter in the solar atmosphere, yielding high-energy gamma rays and neutrinos. Here, we have used the FLUKA simulation code \citep{Ferrari05,Bohlen14} to model cosmic ray propagation and interactions within the solar atmosphere.

\section{Simulation with the FLUKA code}
The FLUKA code is a multi-purpose Monte Carlo code to simulate hadronic and electromagnetic interactions. Details of the modelling of the hadronic interactions treated in FLUKA are discussed in Ref.~\citep{Mazziotta20}. FLUKA has been developed to accurately track both charged and neutral particles in the presence or absence of magnetic fields, even in complicated geometries. The technical details of our simulation using the FLUKA code are discussed in the box.

\begin{figure*}[h!]
\begin{framed}
Our code uses a spherical reference frame centered on the Sun, described as a sphere of radius $R_{\odot} = 6.9551\times 10^{10}$ cm. We have constructed the chromosphere (typical height, h = 2000 Km) above the photosphere that is divided into 100 layers (i.e., shells) with matter (protons) densities following the chromosphere's matter density profile \citep{Seckel91}. The temperature and pressure of the chromosphere are taken from the Ref.~\citep{Mazziotta20}, and the matter density is defined in the layers on an equally spaced logarithmic scale. Below the photosphere, We have implemented 40 layers with densities and pressure according to the standard solar models (SSMs)  \citep{Vinyoles17} for the interior of the Sun below the photosphere, 20 layers from $10^{-7}$ g/cm$^{-3}$ to $10^{-3}$ g/cm$^{-3}$, next 10 layers from $10^{-3}$ to $10^{-1}$ g/cm$^{-3}$, and the inner 10 layers with density greater than $10^{-1}$ g/cm$^{-3}$.

The magnetic field ($B$) close to the Sun is complex and highly time-dependent. The coronal magnetic field is often extrapolated from the measured photospheric magnetic fields. We incorporated the potential field source surface (PFSS) model \citep{Schatten69,Hakamada65} of the solar magnetic field near the Sun in our simulation, in which the field is entirely radial on a sphere of radius $R_{SS}$ (source surface). We have taken the field map for the Carrington Rotation (CR) 2111 (June-July 2011) from the Solar Dynamics Observatory Joint Science Operations Center (JSOC) \citep{Scherrer95}. These field maps are evaluated starting from the different photospheric magnetic field observations \citep{Titov08,Antiochos11,Sun11} with $R_{SS} \simeq 2.5 R_{\odot}$. The magnetic field at the chromosphere is found to be very nearly equal to that at $r\simeq R_{\odot}$, as the height of the chromosphere is $h \ll R_{\odot}$. 

We found that average optical depth of the chromosphere, for varying injection height of a cosmic ray proton above the photosphere, corresponds to that at a height of 330 km above the photosphere. Therefore, we injected accelerated protons with varied kinetic energies from a same height above the photosphere's surface ($r = R_{\odot}$) into the chromosphere uniformly to assess the yields of secondary particles from the Sun. The generation surface corresponds to the outer surface of the outer layer of the chromosphere. The kinetic energies of the primary protons are considered on a grid of 46 equally spaced values on a logarithmic scale ranging from 1GeV to 7 TeV.
\end{framed}
\end{figure*}

\subsection{Gamma-ray and neutrino flux}
The differential intensity of secondary particles (in units of particles eV$^{-1}$cm$^{-2}$s$^{-1}$) emitted from the Sun is given by
\begin{align}
\small
I_{s}(E_{s}) = \int Y_s(E_{s},E_{p}) \frac{dn_p}{dE_{p}}dE_{p}
\end{align}
where, $Y_s$ is represents the yield of secondary particles which can be estimated by counting the secondary particles which escape from the generation surface as defined in Ref.~\citep{Mazziotta20}. We have obtained the yield of the secondary particles (integrated over the solid angle) directly from the FLUKA simulation code \citep{Ahdida22}.

The associated differential flux of high-energy gamma-rays and muon neutrinos reaching the Earth from the Sun may be represented as
\begin{eqnarray}
\frac{d\Phi_{\gamma/\nu_{\mu}}}{dE_{\gamma/\nu}} = \frac{R_{\odot}^2}{R^{2}} I_{\gamma/\nu}(E_{\gamma/\nu})
\end{eqnarray}
where $R$ is the distance of the Sun from Earth.

\subsection{Solar cosmic-ray flux}
The runaway cosmic ray protons diffuse in the interplanetary medium after they escape from the Sun, with a diffusion coefficient that can be represented as $D = D_0(\frac{E_p}{10\;GeV})^{\delta}$, where $D_0 \simeq \frac{1}{3} \lambda_d c$ represents the diffusion coefficient of solar cosmic rays at 10 GeV energy and $\delta$ indicates a constant having values ranging between $0.3$ to $0.7$ \citep{Berezinsky90,Strong07}. Here, $\lambda_d$ denotes the diffusion mean free path, which can have values of ($1-2$) a.u. at $10$ GeV energy at heliocentric distances of $1$ a.u \citep{Pei10}. The expected solar cosmic ray flux reaching the earth after diffusive propagation (considering continuous emission) \citep{Aharonian96} is

\begin{eqnarray}
\small
J_{p} = \frac{c Q_p}{4\pi D R} {erfc} \left({\frac{R}{R_{dif}}} \right)
\label{Eq:2}
\end{eqnarray}
where, $Q_p = 4 \pi R_{\odot}^2 I_p$ represents the luminosity of outgoing protons from Sun, $R_{dif}$ denotes the diffusion radius of cosmic rays, $erfc(z)$ represent the error function of $z$ \citep{Aharonian96}, and $R$ is the distance of the Sun from Earth.

The expression for the resultant maximum amplitude of anisotropy (for $R_{dif} \gg R$) for a continuously emitting source can be obtained by following Refs.~\cite{Osborne76, Bhadra06} and can be written as
\begin{equation}
 \beta_p =  h(E_p) \frac{3D}{c R}
\label{Eq:4}
\end{equation}
where $h(E_p)$ denotes the ratio of the expected cosmic ray flux from Sun to the total observed cosmic ray flux from all sources (i.e. $4\pi$ direction) at same energy energy $E_p$ \cite{Bhadra06}. Here, we use the observed cosmic-ray flux at earth to be $1.8 \left( \frac{E}{GeV} \right)^{-2.7}$ cm$^{-2}$ s$^{-1}$ sr$^{-1}$ GeV$^{-1}$ \cite{Workman22}, where $E$ is the energy of the cosmic rays (including rest mass energy). 

\section{Results}\label{secr}
In a recent study, Mazziotta et al. (2020) \citep{Mazziotta20} showed that the gamma-ray flux generated by the interaction of galactic cosmic rays with the solar atmosphere is sensitive to that observed by Fermi-LAT in GeV energies while taking into account an original PFSS magnetic field map enhanced by a factor of 20 near the Sun to closely follow the BIFROST model \citep{Mazziotta20,Carlsson16,Gudiksen11}. We have carried out our simulation using the same magnetic field setup (PFSS$-$BIFROST model \citep{Mazziotta20}) to obtain the yield of secondary particles escaping from the Sun at different energies of primary solar-originated cosmic-ray protons. Consequently, we estimated the flux of gamma rays, neutrinos, and the solar-originated cosmic rays reaching the earth from the solar disk by following the methodology discussed above.

\begin{figure}[t]
  \begin{center}
  \includegraphics[width = 0.47\textwidth,height = 0.45\textwidth,angle=0]{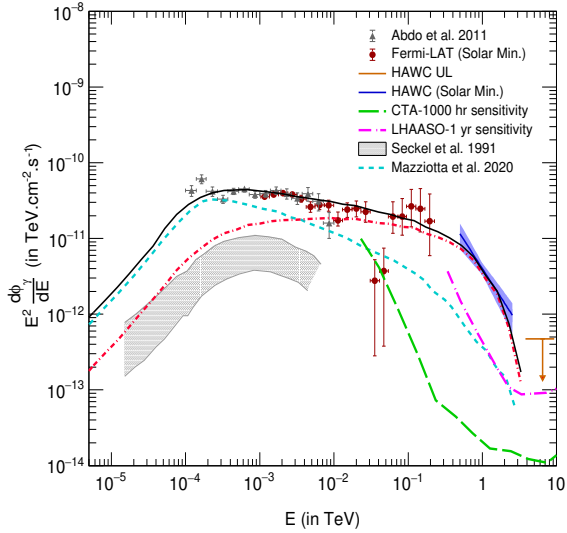}
\end{center}
  \caption{The estimated differential gamma-ray flux reaching Earth from the solar disk during solar minimum. The red dash-dotted line indicates the expected gamma-ray flux from solar disk originating due to the solar cosmic rays according to our model with FLUKA code. The galactic cosmic ray contribution to the gamma-ray flux, as estimated by the Refs.~\citep{Mazziotta20,Seckel91}, are displayed. The black continuous line denote the expected overall gamma-ray flux (including the flux estimated by the Ref.~\citep{Mazziotta20}). The detection sensitivity of the CTA detector for 1000 hrs \citep{Funk13}, and the LHAASO detector for 1 year \citep{Vernetto16} are also shown.}
\label{Fig:1}
\end{figure}

\begin{figure}[t]
  \begin{center}
  \includegraphics[width = 0.47\textwidth,height = 0.45\textwidth,angle=0]{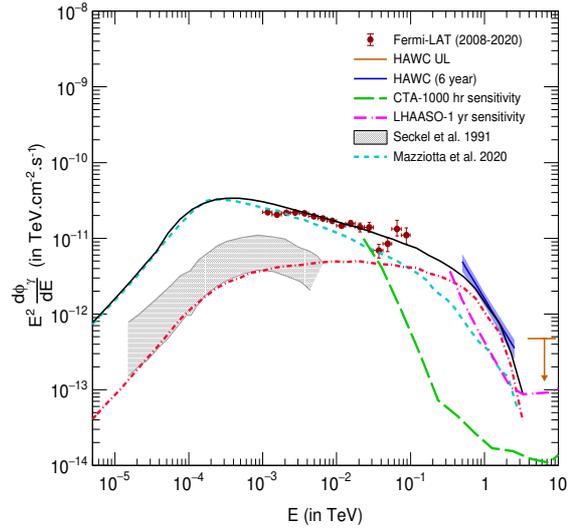}
\end{center}
  \caption{Same as Fig.~\ref{Fig:1}, but for the span of full solar cycle.}
\label{Fig:2}
\end{figure}

The contribution to the observed gamma-ray flux from the solar disk due to the hadronic interaction of the solar cosmic rays with the matter in the solar atmosphere has been computed. To explain the observed gamma-ray spectrum during the solar minimum, we required a production spectrum of solar cosmic rays (protons) with a spectral index of $\alpha_p = 1.95$, and a solar cosmic ray luminosity of $L_p = 2.8\times 10^{19}$ erg/s. On the other hand, a spectral index of $\alpha_p = 1.90$, and a relatively lower cosmic-ray luminosity of $7\times 10^{18}$ erg/s are needed for the entire solar cycle. For both situations, a maximum energy of $E_{p,max} = 7$ TeV for the solar cosmic ray protons is found necessary to match the observed gamma-ray spectrum. Given that the total mechanical flux at the base of the chromosphere for heating the solar atmosphere is of the order of $10^{6}$ erg cm$^{-2}$ s$^{-1}$ \cite{Ulmschneider70}, their power is found to be around $10^{29}$ erg/s. The power of the chromosphere shocks is found to be about $10^{26}$ erg/s (using the shock wave energy flux as given in Ref.~\cite{Kuzma19}), which is a fraction of total power. Thus, the chromosphere shocks have enough power to provide the required luminosity ($\sim 10^{19}$ erg/s) to cosmic ray particles. Thus, the acceleration efficiency of protons in solar atmosphere needed is just $\eta \sim 10^{-7}$. 

The entire gamma-ray emission from the solar disk, including both the solar cosmic ray component (our work) and the galactic cosmic ray contribution \citep{Mazziotta20}, has been found to match the observations consistently. 
The overall gamma-ray emission from the solar disk is due to the solar cosmic ray contribution (our work) plus the galactic cosmic ray contribution \citep{Mazziotta20}. The corresponding gamma-ray fluxes during solar minimum and for the entire solar cycle are shown in Fig.~\ref{Fig:1} and Fig.~\ref{Fig:2} respectively along with the different experimental observations and theoretical expectations. 

Theoretically, the spectral slope due to strong shocks (large Mach number) in diffusive shock acceleration is 2 \citep{Bell}. The non-linear diffusive shock acceleration admits harder spectral slope. The spectral slope for the solar cosmic rays found here is close to the theoretical expectation for strong shocks. We considered that the Sun-accelerated cosmic rays injected all over the chromosphere uniformly. If instead, the Sun-accelerated cosmic rays emerged from the mid-point of the photosphere, we require an even lower flux of solar cosmic rays but we have to take a harder spectrum to off-shoot the over-production of low energy secondaries. The spectral index obtained here, by matching the observed gamma ray spectrum, is found sensitive to magnetic field in chromosphere and the contribution of the galactic cosmic rays to the observed gamma ray spectrum. A better estimate of spectral slope can be made when both gamma rays and neutrino spectra are available.      

Analysis of the Fermi-LAT data reveals a dip in the solar gamma ray flux roughly between 30 GeV to 50 GeV for the cycle 23/24 minimum \citep{Linden18,Linden22} which was totally unanticipated and is not theoretically understood so far. However, such spectral feature is not observed during the most recent cycle 24/25 minimum \citep{Linden22}. So, the possibility of statistical fluctuations as the cause of the dip cannot be completely ruled out at this stage and hence we have not addressed this issue here.

\begin{figure}[t]
  \begin{center}
  \includegraphics[width = 0.47\textwidth,height = 0.4\textwidth,angle=0]{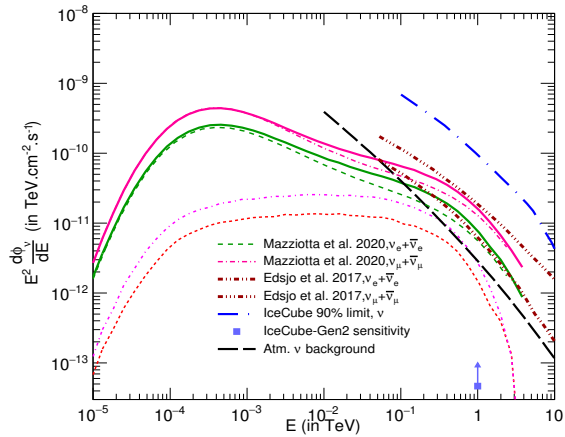}
  \end{center}
  \caption{The estimated differential neutrino flux reaching Earth from the solar disk during solar minimum. The red small-dashed and magenta small-dash-single-dotted lines show the expected electron neutrino ($\nu_e+ \bar{\nu}_e$) and muon neutrino ($\nu_{\mu}+ \bar{\nu}_{\mu}$) flux respectively from the solar disk, originating due to the solar cosmic rays according to our model with FLUKA code. The galactic cosmic ray contribution to neutrinos from the solar disk as estimated by Refs.~\citep{Mazziotta20,Edsjo17}. The green and pink continuous lines indicates the overall electron and muon neutrino flux (including the flux estimated by the Ref.~\citep{Mazziotta20}) from the solar disk, respectively. The IceCube's 90\% limit for neutrinos \citep{Aartsen21_2}, and the expected atmospheric muon neutrino flux \citep{Honda07} from the solid angle in the sky made by the solar disk are shown. The light blue data point at 1 TeV represents the point-like sensitivity of IceCube-Gen2 neutrino telescope during 10 years of observation of a source at $0^{\circ}$ declination assuming an $E^{-2}$ spectrum \citep{Ambrosone21}.}
\label{Fig:3}
\end{figure}

The estimated corresponding neutrino flux is displayed in Fig.~\ref{Fig:3} along with the expected atmospheric muon neutrino flux \citep{Honda07} from the direction of the solar disk. 
It is found that the estimated muon neutrino flux at 1 TeV energy is well within the reach of the sensitivity of IceCube-Gen2 \citep{Aartsen21}, a future extension of the IceCube detector. However, the galactic cosmic rays striking the solar atmosphere generate neutrinos on their way to Earth via hadronic cascade interactions within the Sun \citep{Edsjo17}. These neutrinos might potentially be detected by forthcoming neutrino detectors and will act as a background for searches of neutrinos originating from solar cosmic rays during the quiet stage. 

\begin{figure}[t]
  \begin{center}
  \includegraphics[width = 0.49\textwidth,height = 0.7\textwidth,angle=0]{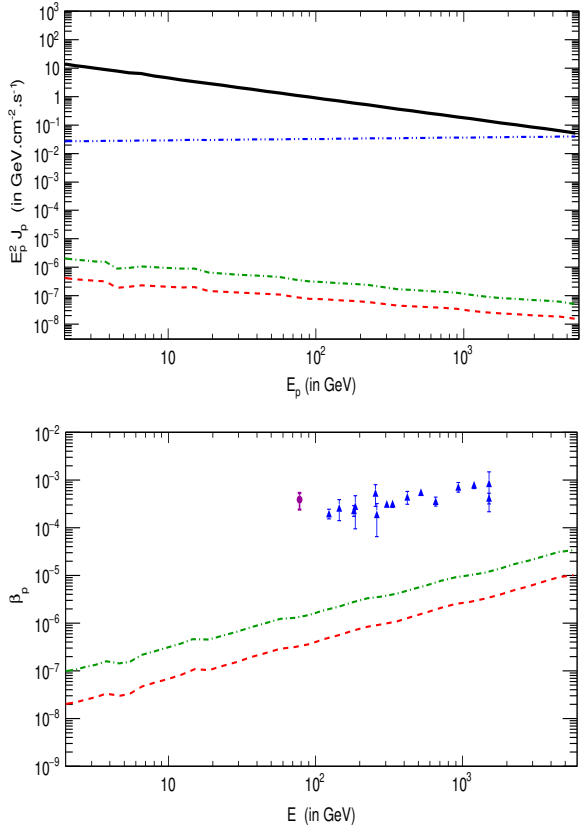}
\end{center}
  \caption{The cosmic ray flux reaching at earth. Top: The black continuous line represents the isotropic observed cosmic ray flux from $4\pi$ direction \citep{Workman22} at earth orbit. The blue dash-double dotted line represents the expected solar cosmic ray production spectrum at Sun. The green dash-single-dotted, and red dashed lines denote the expected cosmic ray flux reaching at earth from solar disk during solar minimum and over the span of full solar cycle, respectively . Bottom: The black continuous, and magenta dashed lines represents the estimated anisotropy of cosmic rays from solar disk during solar minimum and over the span of full solar cycle. The experimental cosmic-ray anisotropy results are taken from: Fermi-LAT measurements \citep{Ajello19} and underground muon (UG-$\mu$) detectors (\cite{Amenomori05} and references therein).}
\label{Fig:4}
\end{figure}

\begin{table*}[ht]
    \begin{minipage}{\linewidth}
        \renewcommand\footnoterule{}
        \renewcommand{\thefootnote}{\alph{footnote}}
  \caption{Relative cosmic ray deficit in the Sun shadow.}\label{table1}
  \begin{tabularx}{\textwidth}{Xccccc}
  \toprule
       $N^{obs}_{def}$    &  Energy  & Solar activity & {$N^{bkg}_{def}$ (CSSS$-$Model)}     & \multicolumn{2}{c}{Expected $N^T_{def}$ when}  \\ \cmidrule{5-6}
       (Observations)     &          &                &    & $R_{dif} \simeq R$   &  $R_{dif} \gg R$       \\     
       \midrule
   $-$(0.5-1.2)\% \citep{HAWC22} &   1 TeV & {During 2014 to 2021\footnotemark[1]} &  $\approx -2\%$   &$-1.9\%$   &  $-1.7\%$      \\ 
(HAWC 2022)&       &           &                &                  &         \\
$-$(3.5-5)\% \citep{Amenomori18}   &   3 TeV  & Minimum        & $\approx -4.4\%$   &$-4.2\%$   &  $-3.1\%$                    \\   
 (Tibet-III 2018)                  &          &                &                    &           &         \\ 
       \bottomrule
  \end{tabularx}
      \vspace{-1.5ex}
  \footnotetext[1]{Maximum of solar cycle 24 to the minimum of solar cycle 25.}%
    \end{minipage}
\end{table*}

The corresponding cosmic ray fluxes reaching the Earth from the direction of the Sun, as well as their anisotropy (maximum) in comparison to the measured isotropic cosmic ray flux at all energies, have been estimated using the standard expressions (see for instance \cite{Bhadra06}) (for $R_{dif} \gg R$) and are shown in Fig.~\ref{Fig:4}. Here, we consider on average $\lambda_d \simeq 1.5$ a.u. at 10 GeV energy and $\delta = 0.5$ for solar cosmic rays in the interplanetary space at heliocentric distance of 1 a.u. \citep{Pei10}. The estimated (maximum) amplitude of anisotropy due to the Sun is found small, within the observational bound. Above $\sim 1$ TeV energies, cosmic ray shadow about the solar disk has been observed by different observatories \citep{Zhu13,Amenomori18,HAWC22}. The relative deficit between the number of events in the shadow region around the Sun, $N_{shadow}$, and the average number of events in the background regions, $\left< N_{bkg}\right>$, is described by $N_{def} = \frac{N_{shadow} - \left< N_{bkg}\right>}{\left< N_{bkg}\right>}$ \citep{HAWC22,Amenomori18}. The simulated intensity deficit in the Sun's shadow in galactic cosmic rays with the current-sheet source-surface (CSSS) model \citep{Zhao95} without considering solar contribution to cosmic rays in TeV energies ($N^{bkg}_{def}$) had been studied earlier \citep{Amenomori18}. Here, we have estimated the relative cosmic ray deficit ($N^{T}_{def}$) considering the solar contribution to overall cosmic ray flux which better matches the observations as shown in Table~\ref{table1}. Here, we adopted a sky window radius of $1.1^{\circ}$ and $0.9^{\circ}$ (angular resolution of respective detectors) to estimate the total cosmic ray deficit in the Sun shadow for HAWC (2022) \citep{HAWC22} and Tibet-III (2018) \citep{Amenomori18} observations respectively. 

It is now important to discuss how to distinguish the case of incoming galactic cosmic ray protons from the case of outgoing solar accelerated protons. When the high energy gamma rays are produced by incoming galactic cosmic ray protons hitting the Sun's surface, the Sun limb should be brighter than the disk (see fig. 10 of Ref. \cite{Mazziotta20}). On the other hand, we would expect that the flux of gamma rays produced by outgoing protons should be more uniform across the Sun disk if the Sun accelerates cosmic rays. Moreover, at high energies, secondaries are mostly created in the forward direction. Fortunately, Fermi-LAT detector has an angular resolution of about 0.1 degrees above 10 GeV energies (the angular radius of the Sun corresponds to 0.264 degrees) \cite{Linden18}. Taking advantage of this, Linden et al. (2022) \cite{Linden22} reported recently that the gamma-ray emission, as observed by Fermi-LAT at energies of $10-50$ GeV, is nearly uniform across the solar surface, which favors the case of outgoing protons (our work). Because the gamma-ray flux from the solar disk is reduced at relatively higher energies, reconstruction of this kind is challenging. A future gamma-ray telescope with better sensitivity and angular resolution at high energies may clearly discriminate the cases of incoming and outgoing protons.

\section{Conclusion}
We conclude that the nature and the flux level of the observed gamma ray emission from the solar disk indicates that the (quiet) Sun is a cosmic ray emitter with maximum energy around a few TeV. We found that the hadronic interaction of solar cosmic rays with solar matter above the photosphere can consistently resolve the puzzle of the origin of the reported gamma rays from the solar disk. The observed cosmic ray flux deficit during solar shadow is found to match better with theoretical estimates when such solar cosmic ray flux is included. Therefore, such an inference is physically not unrealistic and is found consistent with all the concerned observations from the Sun. The upcoming IceCube-Gen2 neutrino detector should be able to test the proposed model by monitoring TeV muon neutrinos from the solar disk. If such evidence for the Sun as a cosmic ray TeVatron is further validated by neutrino observations, the suggested idea should have significant implications in our understanding of the origin of the galactic cosmic rays.


\bmhead{Acknowledgments}
The authors would like to thank two anonymous reviewer for valuable remarks that helped us improve the manuscript. PB thanks the SERB (DST), Government of India for the financial support, under the N-PDF fellowship reference number PDF/2021/001514.


\begin{itemize}
\item Conflict of interest/Competing interests: 
The authors declare no competing interests.
\item Availability of data and materials: This manuscript has no associated data and material.
\item Authors' contributions: The first author performed the analysis part. All authors contributed equally to the rest of this work.
\end{itemize}

\end{document}